# Exploring the Convergence of HCI and Evolving Technologies in Information Systems

Rajan Das Gupta,[1, a)] Ashikur Rahman,[2, b)] Md Imrul Hasan Showmick,[3, c)] Md. Yeasin Rahat,[1, d)] and Md. Jakir Hossen[4, e)]

[1]*American International University-Bangladesh, Bangladesh*
[2]*Northern University Bangladesh, Bangladesh*
[3]*Brac University, Bangladesh*
[4]*Multimedia University, Malaysia*

[a)]18-36304-1@student.aiub.edu
[b)]ashikxql@gmail.com
[c)]imrul.hasan.showmick@gmail.com
[d)]20-43097-1@student.aiub.edu
[e)]Corresponding author: jakir.hossen@mmu.edu.my

**Abstract.** Modern technology driven information systems are part of our daily lives. However, this deep integration poses new challenges to the human computer interaction (HCI) professionals. With the rapid growth of mobile and cloud computing and the Internet of Things (IoT), the demand for HCI specialists to design user-friendly and adaptable interfaces has never been more pressing. Especially for diverse user groups such as children, the elderly and people with disabilities who need interfaces tailored to their needs regardless of time and location. This study reviewed 50 recent papers on HCI interface design for modern information systems. The goal is to see how well these methods address the demands of current technology. The findings show that most HCI design methods are still based on old desktop models and do not support mobile users and location-based services well. Most existing interface design guidelines do not align with the flexibility and dynamism of emerging technologies. The goal of this study is to improve interface design by combining agile methodologies with human-centered design principles. Future studies should also incorporate both qualitative and quantitative approaches, particularly in the context of cloud-based technologies and organizational information systems. This approach aims to bridge the gap between current interface design practices and the changing technological landscape.

## INTRODUCTION

The rapid expansion of Information Technology (IT) continues to reshape how societies and organizations operate [1]. Emerging technologies are now embedded in everyday life, influencing not only personal routines but also business models [2], global platforms such as crowdsourcing [3], and decision-making infrastructures across domains [4]. These transformations have fueled ongoing academic debates on the nature of technological change [5] and raised urgent questions about how technology shapes collective behavior, values, and social systems [6]. Within this shifting landscape, Human–Computer Interaction (HCI) plays a critical role in ensuring that interactive systems remain usable, accessible, and meaningful [7]. Yet, despite decades of advances, designing effective user interfaces continues to be a persistent challenge. The emergence of technologies such as cloud computing [8], virtualization [9], mobile platforms [10], and the Internet of Things (IoT) [11] has introduced novel design constraints that traditional methods often fail to address. For example, mobile devices—now used globally by children, the elderly, and people with disabilities—[12, 13, 14] are still predominantly designed with desktop-oriented paradigms [15], which neglect the contextual and situational needs of mobile users [10].

Similar issues arise in IoT and decision-support applications in healthcare and agriculture, where complex environments demand adaptive and user-centered interaction models [16, 17, 18]. At the same time, new paradigms such as gamification [19] and Big Data–driven feedback systems [20] hold promise but lack rigorous evaluation regarding their impact on usability, engagement, and trust [21]. As a result, despite significant technological progress, the field still faces a critical gap: how can HCI principles be effectively integrated into rapidly evolving digital ecosystems?

This study systematically reviews existing literature to evaluate how HCI professionals design user interfaces, highlighting current opportunities, limitations, recurring challenges, and future directions for advancing HCI in complex and evolving technological contexts.

# BACKGROUND & RELATED WORK

The following sections synthesize prior research on information system design and human–computer interaction techniques, providing a critical analysis that not only consolidates existing knowledge but also exposes gaps and limitations. This integrated perspective establishes the foundation for advancing more adaptive, user-centered, and context-sensitive design approaches in contemporary digital ecosystems.

## Information System Design

Designing modern information systems has become increasingly complex due to rapid technological change and evolving user expectations [5]. Beyond technical performance, systems must now satisfy usability, accessibility, and adaptability requirements in environments that are more interconnected and dynamic than ever before. User involvement has emerged as a cornerstone of successful system design. Engaging stakeholders throughout the process not only improves satisfaction but also ensures that systems align with real-world practices rather than idealized models [22, 23]. This is particularly important as mobile and cloud technologies expand access for populations such as the elderly and people with disabilities, yet simultaneously introduce new design challenges related to context-awareness, interoperability, and security [10, 14].

Organizations increasingly demand interfaces that are flexible, engaging, and aligned with both organizational goals and user needs [24]. Yet mobile interfaces still lack unified standards and diverge from desktop paradigms, underscoring the need for frameworks that balance usability, aesthetics, and efficiency across platforms. HCI addresses this challenge by emphasizing user-centeredness, reducing cognitive load, and positioning the interface as the "face" of the system, where design, interaction, and multimodal input converge to shape user experience.

A critical weakness of many existing design methods is their limited grounding in real-world contexts. Systems in domains such as healthcare, for example, operate in environments characterized by uncertainty, variability, and high user diversity. Designing for such complexity requires approaches that go beyond abstract

**TABLE 1.** Key concepts in HCI and information systems design

| Title | Methods | Core Concept |
|---|---|---|
| Adaptive UI | Survey | UI design must adapt to different users and organizational needs. |
| Health Hackathons | Hierarchical coding system | Involving stakeholders in system design improves effectiveness. |
| IoT Healthcare | One-shot survey | Emerging technologies enhance quality of life and safety. |
| Mobile Usability | Case study | Mobile UI design differs from desktop paradigms and lacks universal standards. |
| Aging Design | Experimental | Participatory design is essential for creating inclusive digital technologies. |
| HCI Development | Manual literature coding | Specific HCI approaches improve interactive system development. |
| Electronics Learning | Survey | UI design should cater to diverse user needs, including mobile-specific approaches. |
| Art & HCI | Hierarchical coding system | Collaboration between art, designers, and HCI enhances user experiences. |
| Accounting Systems | One-shot survey | User involvement is crucial in designing effective accounting systems. |
| UX Evolution | Case study | UX methodologies are evolving beyond human-centered approaches. |
| SE & HCI | Experimental | Integrating SE and HCI models enhances interactive system re-engineering. |
| Agile UCD | Manual literature coding | User-centered agile development principles provide a structured approach. |
| IS Failure | Survey | Poor UI design contributes to information system failures. |

## Review of HCI Design Approaches

Integrating human–computer interaction (HCI) with software engineering is critical for redesigning complex systems and closing communication gaps. Agile and user-centered approaches embed usability into development, while poor interfaces remain a major cause of system failure [25, 26]. Lean and Agile methods address these risks through iteration, feedback, and user involvement, offering adaptable design practices [27]. Table 1 summarizes key studies and methods shaping current understandings of HCI and information system design.

# METHODOLOGY

## Research Design

This study adopts a systematic literature review (SLR) methodology, which is widely recognized for providing a structured and transparent approach to synthesizing existing knowledge. Building on prior work [28], the review focuses on examining human–computer interaction (HCI) approaches to information system development and user interface (UI) design principles. The SLR design enables the identification of recurring challenges, evaluation of methodological trends, and synthesis of insights that inform both theory and practice in contemporary HCI and information systems research.

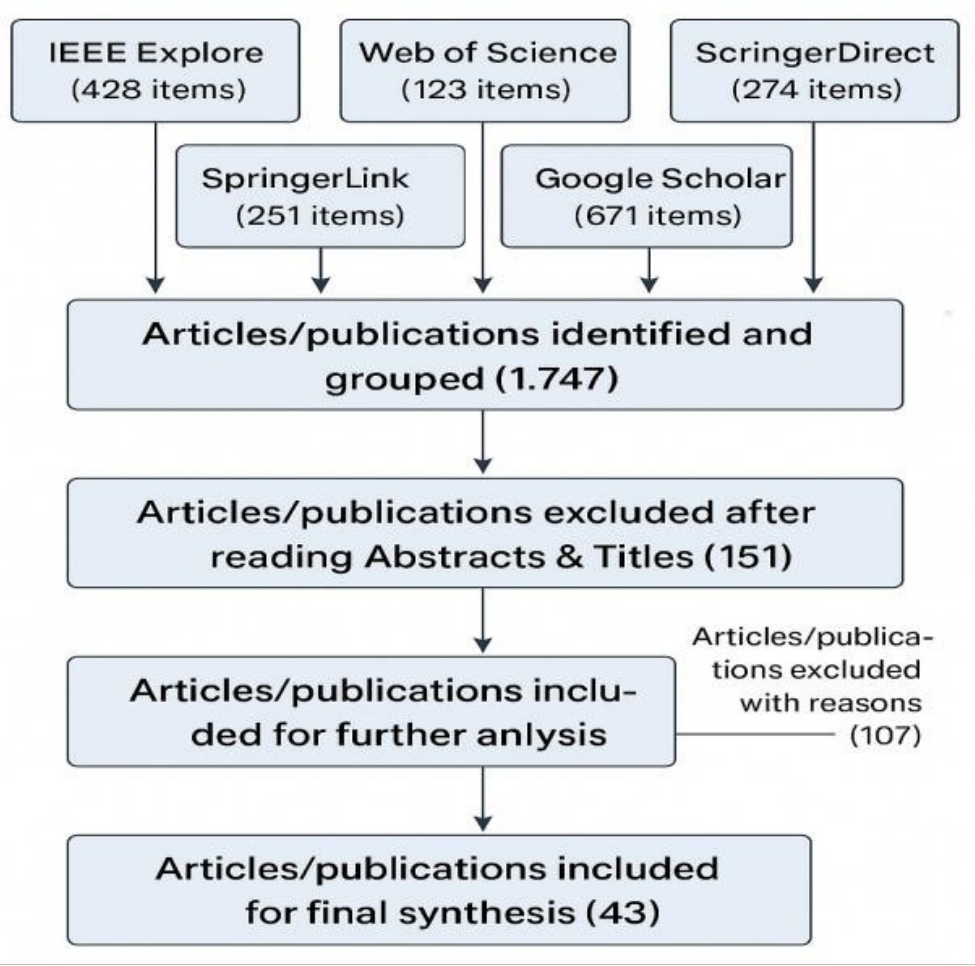

**FIGURE 1.** Systematic literature review flowchart for article selection and synthesis

## Database Search

After defining the scope of the review, a structured search for relevant studies was conducted across leading scholarly databases. Specifically, **IEEE Xplore**, **Web of Science**, **ScienceDirect**, **SpringerLink**, and **Google Scholar** were queried to identify articles related to human–computer interaction approaches, information system development, and user interface design principles. These databases were selected for their comprehensive coverage of computer science, engineering, and interdisciplinary research, ensuring both breadth and depth in the literature review process.

## Search Strategy

The search process began with the use of carefully selected keywords that capture the scope of this review. Terms included: ***"human–computer interaction approaches"***, ***"HCI design paradigm"***, ***"user interface (UI) design"***, ***"information systems development"***, ***"systems design methodologies"***, ***"Internet of Things (IoT)"***, and ***"mobile plat-*** ***forms"***. These keywords were applied across multiple scholarly databases to ensure broad coverage of relevant literature. A substantial number of articles were retrieved, screened, and filtered for relevance, with the final set of studies selected for detailed analysis. The overall process is illustrated in Figure 1.

## Selection criteria

As seen from Figure 1, searched outcomes from the databases produced (1,747 articles), which were group together and exported to excel for screening. There's no time limit with regards to articles publication period. In the second stage of our selection process, 1,513 articles were excluded after reading their abstracts and titles, as they were found to have no relevance to the study. Furthermore, articles written in other languages, other than English language, with inadequate details on the role of HCI in information systems design, so much emphasis on engineering aspect of HCI, or focused more on information systems activities without considering design problems were also excluded. The remaining articles were included for further screening, after which 84 articles were excluded either due to lack of full access or duplication. Finally, 43 articles were fully read and analyzed in the review.

## RESULT & DISCUSSION

This study examined how Human–Computer Interaction (HCI) methods are applied to modern technologies such as mobile platforms, cloud systems, and IoT in domains like healthcare and agriculture. Despite rapid technological change, many approaches remain similar to those used two decades ago, emphasizing structure, visuals, and interaction. These traditional methods struggle to meet the complexity of contemporary systems, which must function across distributed, dynamic, and highly personalized contexts. Established HCI principles—task orientation, consistency, and user-centered needs—remain valuable but are still rooted in desktop paradigms. This limits their suitability for mobile and cloud-based applications, which are accessed ubiquitously by diverse groups, including the elderly, young, and individuals with disabilities. Such reliance highlights the need for adaptive strategies that align more closely with real-world use.

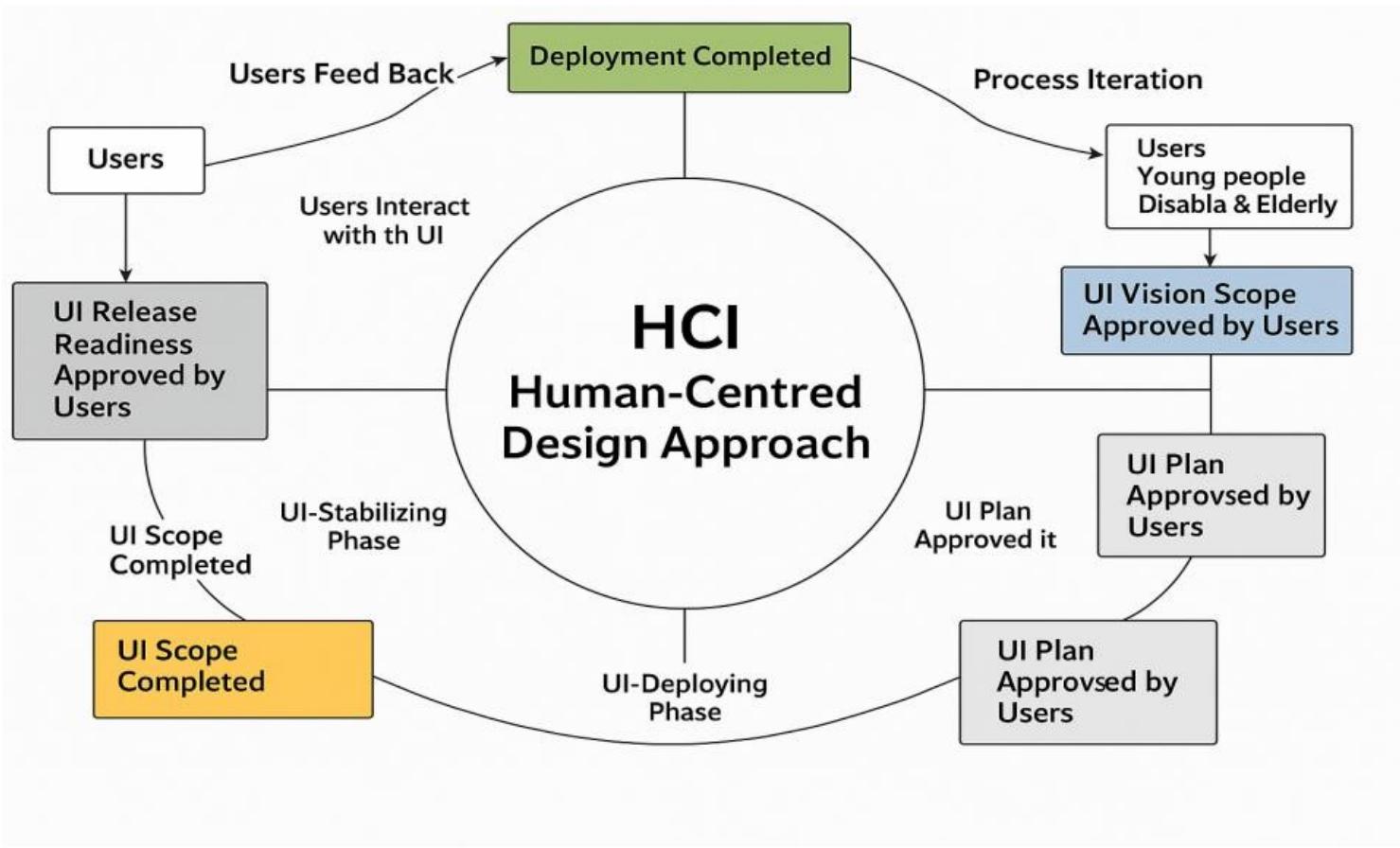

**FIGURE 2.** Study proposed integrated user-interface design methodology

Conventional User-Centered Design (UCD) alone is no longer sufficient. As the field shifts toward Human-Centered Design (HCD) and User Experience (UX), the focus must extend beyond usability to include affective, social, and contextual dimensions of interaction. To address these challenges, this study recommends combining HCD with Agile development methods. Agile's iterative cycles enable incremental development, frequent user feedback, and continuous refinement. This approach ensures that users remain engaged throughout—from design to deployment—producing interfaces that are functional, inclusive, and meaningful. As illustrated in Figure 2, each interface version can be validated directly with users, creating adaptive systems that meet the demands of contemporary digital ecosystems.

# CONCLUSION

This study reviewed 43 papers, revealing both a limited body of literature and ongoing debates about the positioning of HCI within information systems research. Many contributions leaned heavily on engineering-focused design, often overlooking broader IS perspectives. The findings indicate that current HCI design approaches are largely inadequate for modern systems, as they remain rooted in static desktop paradigms that fail to support the dynamic, location-based requirements of mobile, cloud, and IoT technologies. The shift from human-centered to user-experience design underscores how emerging technologies—such as smart home systems, gamification, and IoT healthcare—demand more interactive, responsive, and visually adaptive interfaces. Even user-centered approaches were found insufficient to address the complexity of today's UI challenges. To overcome these gaps, this study recommends integrating Human-Centered Design with Agile development, enabling iterative feedback and user involvement throughout the design process.

While this review focused primarily on mobile platforms, future research should extend to cloud-based and distributed systems, employing more empirical methods to assess how HCI can evolve to meet the demands of increasingly complex and socially impactful technologies.

# ACKNOWLEDGMENT

The authors would like to thank **Multimedia University** and the **ELITE Research Lab** for their support of this research.